# Number of aftershocks in epidemic-type seismicity models

## G. Molchan

23 April 2024

Institute of Earthquake Prediction Theory and Mathematical Geophysics,
Russian Academy of Science, 84/32 Profsoyuznaya st.,117997, Moscow,RF

Email: molchan@mitp.ru

## Abstract

Let $V_M(m_\bullet)$ be the number of $m \geq M$ aftershocks caused by $m_\bullet$ event. We consider the $V_M(m_\bullet)$ distribution within epidemic-type seismicity models, ETAS(F). These models include the Gutenberg-Richter law for magnitude and Utsu law for average $m_\bullet$- productivity, but differ in the type of $F$ distribution for the number $\nu(m_\bullet)$ of direct aftershocks. The class of $F$ is quite broad and includes both the Poisson distribution, which is the basis for the regular ETAS model, and its possible alternative, the Geometric distribution. Instead of the traditional threshold $M = m_\bullet - \Delta$ we consider $M = m_a - \Delta$, where $m_a$ is the distribution mode of the strongest aftershock. Under these conditions we find the limit $V_M(m_\bullet)$ distribution at $m_\bullet \gg 1$. In the subcritical case, the limit distribution is extremely simple and identical to the $\nu(m_\Delta)$ distribution with a suitable magnitude $m_\Delta$. Theoretical results of this kind are lacking even for the regular ETAS model. Our results provide an additional opportunity to test the type of F-distribution, Bath's law, and the very concept of epidemic-type clustering.

## 1.Introduction

We are interested in the distribution of the number of aftershocks $m \geq M$, $V_M(m_\bullet)$, caused by an $m_\bullet$ event. To weaken the dependence of the distribution on $m_\bullet$, the so-called $\Delta$-analysis is traditionally used, assuming that M is equal to $M_\bullet = m_\bullet - \Delta \geq m_c$. Such approach does not give a clear answer about the $V_{M_\bullet}(m_\bullet)$ distribution.

Solovyev and Solovyeva [1962] were probably the first to postulate the Geometric distribution for $V_{M\bullet}(m_\bullet)$. This conclusion was based on relatively scarce data (1954-1961) for the Pacific Belt ($m_\bullet \approx 7; \Delta = 2$) and for the Kamchatka and Kuril Islands region ($m_\bullet \approx 6; \Delta = 2$). Shebalin et al. [2018] confirmed the hypothesis using 850 aftershock sequences with main shocks $m_\bullet \geq 6.5, \Delta = 2$ from the ANSS (1975-2018) catalog.

In the same time, Kagan [2010], using PDE (1977-2007) catalog, $m_\bullet = 7.1 - 7.2, \Delta = 2.5$, show a bimodal $V_{M\bullet}(m_\bullet)$ distribution with a dominant peak at zero value of $V_{M\bullet}(m_\bullet)$.

Zaliapin and Ben-Zion (2016) used the global catalog (NCEDC,1975-2015, $m_c = 4, \Delta = 2$, 2000 aftershock sequences) and clusters triggered not necessarily by the main (strongest) event. It was shown that the survival function $P(V_{M\bullet}(m_\bullet) > n)$ looks like a power-law $Cn^{-2}$ for locations with high heat flow, and like a log-normal distribution otherwise.

Molchan et al (2023) used epidemic-type cluster models to theoretically analyze the problem and found conditions for the existence of heavy and light tails in the $V_{M\bullet}(m_\bullet)$ distribution. It was also shown that the distribution of $V_{M\bullet}(m_\bullet)$ and the distribution of the number $\nu(m_\bullet)$ of direct aftershocks cannot be of the same type, say Poisson or Geometric (the exact definition of the type will be given later). This statement is inconsistent with the observations of Shebalin et al. [2018, 2020] on the geometric distribution of both $V_{M\bullet}(m_\bullet)$ and $\nu_\Delta(m_\bullet)$, where $\nu_\Delta(m_\bullet)$ is the number of direct aftershocks with $m \geq m_\bullet - \Delta$.

The uncertainty of the $V_{M\bullet}(m_\bullet)$ distribution requires analysis. If the magnitude gap $\delta m_\bullet$ between the main event $m_\bullet$ and the strongest aftershock for example grows with $m_\bullet$, the real range of aftershocks, $\Delta - \delta m_\bullet$, in $V_{M\bullet}(m_\bullet)$ statistics decreases. In such case the choice $M = m_\bullet - \Delta$ excludes the independence of the $V_{M\bullet}(m_\bullet)$ distribution on $m_\bullet$. In particular, $V_{M\bullet}(m_\bullet)$ statistics can be abnormally small for large $m_\bullet$, if the gap $\delta m_\bullet > \Delta$.

It is more flexible way to use $M = m_a - \Delta := M_a$, where $m_a$ is, for example, the peak value in the distribution of the strongest aftershock in the cluster. We used this idea to find the distribution of $V_{M_a}(m_\bullet)$ at $m_\bullet \gg 1$ in the framework of epidemic-type seismicity models, ETAS(F). These models include the Gutenberg-Richter magnitude distribution and the exponential Utsu productivity law, $\lambda(m) := E\nu(m)$. They differ in the type of F distribution for the number $\nu(m)$ of direct aftershocks. The F-class includes both the traditional Poisson distribution and its possible alternative, the Geometric distribution. The magnitude distributions of the strongest aftershock for this class have been studied previously in [Molchan &Varini, 2023].

We are going to show that at the modified choice of $M$, the limit $V_{Ma}(m_\bullet)$ distribution exists and is determined by the distribution of direct aftershocks. Moreover, in the subcritical regime, the limit $V_{Ma}(m_\bullet)$ distribution is identical to the $\nu(m_\Delta)$ distribution with a suitable initial magnitude $m_\Delta$. To the best of our knowledge, theoretical results of this kind are lacking even for the regular ETAS model.

## 2. Preliminary

This paper is a continuation of [Molchan et al, 2022] and [Molchan, Varini, 2024]. The necessary facts from these papers are summarized below.

*2.1 ETAS(F) model.*

The regular epidemic-type aftershock sequence model (ETAS) is defined by means of the conditional intensity of an event at a point of the phase space S= (time, location, magnitude), given the past of the seismic process up to the current time [Ogata,1988]. The model is not flexible enough to describe direct aftershocks [Shebalin et al, 2020]. This shortcoming can be easily overcome by describing a cluster of seismic events in S as a Galton-Watson tree. In the future, the coordinates (time, place) are irrelevant. Therefore, we remind the model in projection to the magnitude.

The initial event of magnitude $m$ generates a random number $\nu(m)$ of direct aftershocks according to the *off-spring law F*:

$$P(\nu(m) = k) = p_k(m). \tag{1}$$

Each of these events is assigned an independent magnitude in accordance with the law $F_1$:

$$P(\mu \leq x) = \int_0^x f_1(m)dm = F_1(x). \tag{2}$$

(Here and hereinafter the threshold magnitude $m_c$ is taken as a reference point, $m_c = 0$).

Each event of the first generation independently generates new ones, following the described rule for its parent, etc. The process continues indefinitely. It stops with probability 1 if the *criticality index*

$$n = \int_0 \lambda(m) f_1(m) dm \leq 1, \tag{3}$$

where $\lambda(m) = E\nu(m)$ is the average *productivity* of the *m*-event. The values $n < 1$ and $n = 1$ correspond to the *subcritical* and *critical* regimes, respectively.

We get the standard ETAS model if F has a Poisson distribution (P), that is

$$p_k(m) = \lambda^k(m) e^{-\lambda(m)} / k! \ . \tag{4}$$

Shebalin et al [2020] found that the Geometric distribution (G),

$$p_k(m) = p^k(m)(1 - p(m)), \ p(m) = \lambda(m)/(1 + \lambda(m)) \ , \tag{5}$$

agrees better with empirical data.

**2.2 *Random Thinning* (RT) *property*.** In the ETAS(F) model, F is a family of distributions parameterized by the magnitude $m$ of the initial cluster event or its productivity $\lambda(m)$, shortly $F = \{F_\lambda\}$, where $\lambda$ is the average value. Below we will consider a special class of $F$ distributions that have the so-called Random Thinning (RT) property. To explain this, let the random variable $v^{(\lambda)} \geq 0$ with mean be the number of elements in some sample and $R_p v^{(\lambda)}$ be the number of those elements that will remain in the sample after random selection of each of them with probability $p$, i.e. $R_p v^{(\lambda)} = \varepsilon_1 + ... + \varepsilon_{v^{(\lambda)}}$, where $\varepsilon_i \in \{0,1\}$ are independent and $P(\varepsilon_i = 1) = p$. For example, the number $v_\Delta(m_\bullet)$ of direct aftershocks with magnitudes $m \geq m_\bullet - \Delta$ is stochastically equivalent to a random variable resulting from the thinning operation of $v(m_\bullet)$ with probability $p = 1 - F_1(m_\bullet - \Delta)$.

We will say that the family of random variables $v^{(\lambda)}, \lambda \geq 0$ has RT property if $v^{(\lambda p)}$ and $R_p v^{(\lambda)}$ are stochastically identical for any fixed $\lambda > 0$ and $0 < p < 1$:

$$R_p v^{(\lambda)} =_{law} v^{(\lambda p)},$$

that is both random variables have the same distributions. In this case, $v^{(\lambda)}$ and $R_p v^{(\lambda)}$ differ only in their mean values, but do not change their type, which corresponds to the reference $v^{(\lambda)}$ distribution with mean $\lambda = 1$.

Since $Ez^{R_p v^{(\lambda)}} = E(1 - p + pz)^{v^{(\lambda)}}$, the generating function of $v^{(\lambda)}$ with RT property has the following representation

$$\varphi(z) = Ez^{v^{(\lambda)}} = \phi(\lambda(z - 1)). \tag{6}$$

The $\phi(w), w < 0$ function uniquely determines the common *distribution type* for the family of random variables $v^{(\lambda)}$. Therefore, below the symbol $\phi_F$ below is used as the F-distribution type.

The class of random variables with the RT property is quite wide. The following statement shows that all types of RT distributions can be obtained as limit points for distributions of $R_{1/\lambda} v^{(\lambda)}$ variables for which $E v^{(\lambda)} = \lambda \gg 1$.

**Statement 1.** . Let $\{\varphi_n(z)\}$ be a sequence of generating functions with means $\lambda_n \to \infty$. Then there exists a subsequence $\{\tilde{n}\}$ such that for any $\lambda > 0$ there exists the limit

$$\lim_{\tilde{n}\to\infty} \varphi_{\tilde{n}}(1+\lambda(z-1)/\lambda_n) = \phi(\lambda(z-1)), |z|<1.$$

where $\phi(z-1)$ is the generating function of some distribution independent of $\lambda$ and having unit mean.

The proof and properties of $\phi(w)$ are described in Appendix.

**Examples.** Let us note important examples of distributions with the RT property:

1) The Negative Binomial distribution F= $NB(\tau)$. This $\tau$ – family of distributions includes the following RT types:

$$\phi_\tau(w) = (1-w/\tau)^{-\tau}, \tau > 0, \quad \phi_\infty(z) = \exp(w) \tag{7}$$

The case $\tau = 1$ corresponds to the Geometric distribution, while the limiting case $\tau = \infty$ corresponds to the Poisson distribution.

2) The Kovchegov - Zaliapin (2021) distribution having the following RT type:

$$\phi_\tau(w) = 1 + w + (-w)^\tau / \tau, 1 < \tau < 2 \tag{8}$$

Seismo-statistical applications of the distributions related to (8) are discussed in [Kovchegov et al., 2022]

The distributions related to RT types (7, 8) have light tails: $p_n \approx cn^{\tau-1}(1+\tau/\lambda)^{-n}, n >> 1$ in (7), and heavy tails $p_n \approx cn^{-\tau-1}, n >> 1$ in (8).

**2.3 Cluster thinning.** In applications, the ETAS(F) model depends on a lower magnitude threshold $m_0$, which we have taken as a reference point. Natural candidates for $m_0$ may be a $m_c$ threshold for the representativeness of magnitudes or a fertility threshold, $m_f$, that determines the ability of the *m*- event to generate aftershocks (personal communication D.Sornette).

The RT property for $F$ distribution allows us to establish a simple relationship between models with different values of $m_0$.

Within an ETAS(F) cluster with $m_0 = 0$, we can identify a sub cluster of causally related events with $m \geq m_f > 0$, starting from the initial event. Following Duquesne and Winkel (2007), we remove from the original cluster all events with $m \leq m_f$ along with their descendants of all generations. Then, we shift the magnitude to use the scale $\tilde{m} = m - m_f$. The resulting sub cluster we will call ETAS(F, $m_f$). The following statement is a simple consequence of the RT property of F

**Statement 2.**

If the ETAS(F) model is such that F distribution has the RT property,

$$E_F \nu(m) = \lambda(m) \quad \text{and} \quad P(\mu > m) = \overline{F}_1(m),$$

then the ETAS(F, $m_f$) model is stochastically equivalent to the ETAS($\tilde{F}$) model, where $\tilde{F}$ has the same RT type as F,

$$E_{\tilde{F}} \nu(\tilde{m}) = \lambda(\tilde{m} + m_f) \overline{F}_1(m_f) \text{ and } P(\mu > \tilde{m}) = \overline{F}_1(\tilde{m} + m_f) / \overline{F}_1(m_f).$$

In the regular case: $\lambda(m) = \lambda_0 e^{\alpha m}$, $\overline{F}_1 = e^{-\beta m}$, the same ETAS($\tilde{F}$) characteristics look similar except for $\lambda_0$, which needs to be replaced with $\tilde{\lambda}_0 = \lambda_0 e^{-(\beta-\alpha)m_f}$. The RT type of F distribution, say Poisson or Geometric, will remain unchanged

As we can see, cluster modeling with different fertility thresholds does not take us beyond the chosen ETAS(F) model with fixed RT type of F

**2.4 Cluster types**. The initial cluster magnitude $m_\bullet$ may be the strongest (dominant) or arbitrary. Therefore, we will consider clusters of both types: $DM(m_\bullet)$ with a Dominant initial Magnitude and $AM(m_\bullet)$ with an Arbitrary initial Magnitude**.** Statement 6 in the Appendix show that any $DM(m_\bullet)$ cluster in ETAS(F) model can be viewed as an $AM(m_\bullet)$ cluster in the ETAS($\hat{F}$) model with a suitable off-spring distribution $\hat{F}$.

**2.5 The strongest cluster event.** The following statement describes the limit distributions of strongest aftershock magnitude in a cluster caused by large event $m_\bullet \gg 1$.

**Statement 3** [Molchan, Varini, 2024]. Let us consider the ETAS(F) model with the distribution $F$ having the RT property, i.e.,

$$E z^{\nu(m)} = \phi_F(\lambda(m)(z-1)), \lambda(m) = E\nu(m).$$

Suppose that

a) $P(\nu(m) = 0) \to 0, m \to \infty$, b) $E\nu^2(m) < \infty$, (9)

c) $\lambda(m) = \lambda_0 e^{\alpha \cdot m}, m \geq 0$, (10)

d) $f_1(m) = \beta e^{-\beta m}/(1 - e^{-\beta M_1}), 0 \leq m \leq M_1 = K_1 m_\bullet$, (11)

where $\alpha \leq \beta$, $1 \leq K_1 < \infty$; in addition, $K_1 \leq \infty$ if $\alpha < \beta$.

Let $\mu_a$ be a maximal magnitude in the AM/DM cluster and $n$ is the criticality index.

**A.** Under $m_\bullet \gg 1$ condition, the following regression relationships are valid for AM/DM clusters:

$$\mu_a = m_a + \rho\varsigma,$$ (12)

where

$$m_a = \alpha\beta^{-1} m_\bullet + \beta^{-1} \ln(\lambda_0/(1-n)), \quad \rho = 1/\beta \quad \text{if} \quad (\alpha < \beta, n < 1);$$ (13)

$$m_a = m_\bullet + \beta^{-1} \ln(\lambda_0/(1 - n/K_1)), \quad \rho = 1/\beta \quad \text{if} \quad (\alpha = \beta, n \leq 1)$$ (14)

(since $n$ is fixed, $\lambda_0$ in (14) is a function of $m_\bullet$: $\lambda_0 = n \cdot (\beta K_1)^{-1}/m_\bullet$);

$$m_a = 2\alpha\beta^{-1}m_\bullet - \beta^{-1}\ln m_\bullet \cdot 1_{\beta=2\alpha} + \beta^{-1}\ln A, \quad \rho = 2/\beta \quad \text{if} \quad (2\alpha \leq \beta, n=1), \tag{15}$$

and $A = 2(\beta\phi_F''(0))^{-1}(\beta - 2\alpha + 1_{\beta=2\alpha})$.

The random component $\varsigma$ in (12) has the limit distribution:

$$P(\varsigma < x) = \phi_F(-e^{-x}), |x| < \infty \tag{16}$$

**B.** In the case $(\alpha < \beta < 2\alpha, n=1)$, the regression relationship (12) depends on the cluster type.

Let $\psi(x), x > 0$ be defined by the following equation

$$\psi + (\beta/\alpha - 1)\psi^{\beta/\alpha}\int_0^\psi u^{-\beta/\alpha-1}[\phi_F(-u) - 1 + u]du = x. \tag{17}$$

Then regression (12) is valid for the **AM**- cluster with the parameters:

$$m_a = m_\bullet + \alpha^{-1}\ln\psi(\lambda_0), \quad \rho = 1/\alpha \tag{18}$$

In the case of a **DM** cluster, the limit $\mu_a$ distribution is as follows

$$P(\mu_a - m_\bullet < -x) \approx \phi_F(-\psi(\lambda_0(1 - e^{-\beta x}))e^{\alpha x}), x \geq 0 \tag{19}$$

For small $\lambda_0 \approx 1 - \alpha/\beta$, (19) admits the following simplification:

$$P(\mu_a - m_\bullet < -x) \approx \phi_F(-\lambda(x)F_1(x)), \quad x \geq 0.$$

*Remark*. The non-random component $m_a$ in the regression relation (12), will be conventionally called *the peak value* of $\mu_a$. In the most interesting case, when F is the Negative- Binomial distribution $NB(\tau)$, this conventional name becomes precise, [Molchan, Varini, 2024].

### 3. Results.

**3.1 *The main equation***, From the structure of the ETAS(F) model, it follows that an *m*-event generates $\{\mu_i, i=1,...,\nu(m)\}$ direct aftershocks, which in turn independently generate clusters of events with magnitude m>M in the size $\{V_M(\mu_i)\}$. All random variables $V_M(\mu_i)$ are independent and equally distributed, given the random nature of the initial event $\mu_i$. As a result we get the following stochastic equation:

$$V_M(m) =_d \sum_{i=1}^{\nu(m)}[V_M(\mu_i) + 1_{\mu_i > M}], \tag{20}$$

where the $1_{\mu_i > M}$ term accounts for the $\mu_i$ aftershock if $\mu_i > M$.

*3.2 Choice of M*.

If $n < 1$, (20) entails

$$EV_M(m_\bullet) = \lambda(m_\bullet)\overline{F}_1(M)/(1-n), \quad \overline{F}_1(M) = \int_M f_1(m)dm. \tag{21}$$

For the exponential $\lambda(m)$ and $f_1(m)$ characteristics presented in (10,11),

$$EV_M(m_\bullet) \approx e^{\alpha m_\bullet - \beta M} \lambda_0 /(1-n) , \quad M_1 \gg 1.$$

Given $M = m_\bullet - \Delta$ we have $EV_M(m_\bullet) \approx Ce^{-(\beta-\alpha)m_\bullet} e^{\beta\Delta} = o(1)$. Hence, to have a nontrivial limit $V_M(m_\bullet)$ distribution at $m_\bullet \gg 1$, the choice of $M$ should be such that $0 < c < EV_M(m_\bullet) < C, m_\bullet \to \infty$, that is $M = (\alpha/\beta)m_\bullet + const$. The non-random component $m_a$ of the strongest aftershock (see Statement 3) has exactly the same form in the subcritical regime.

The choice $M = m_a$ is quite probable and in the general case. Indeed the events $\{V_M(m_\bullet) = 0\}$ and $\{\mu_a < M\}$ are identical for any $M$, and hence

$$P\{\mu_a < M\} = P(V_M(m_\bullet) = 0) = Ez^{V_M(m_\bullet)}\big|_{z=0} \tag{22}$$

Assuming $M = m_a$ the right part of (22) will have a nontrivial limit at $m_\bullet \gg 1$.

Ratio (22) shows that the problem under consideration includes the problem of distribution of the strongest aftershock.

### 3.3 The limit $V_{Ma}(m_\bullet)$ distribution

We will consider the ETAS(F) model described in Statement 3 and use the notation adopted there.

**Statement 4**

Under the assumptions of Statement 3, the $(V_{Ma}(m_\bullet), M_a = m_a - \Delta)$ statistics has a non-trivial limit distribution at $m_\bullet \gg 1$.

**I.** In $(\alpha < \beta, n < 1)$ and $(\alpha = \beta, n \leq 1)$ regimes, the limit $V_{Ma}(m_\bullet)$ distribution is independent of the AM/DM cluster type and has generation function $\phi_F(e^{\beta\Delta}(z-1))$ corresponding to the $v(m_\Delta)$ statistics with mean $\lambda(m_\Delta) = e^{\beta\Delta}$, that is

$$(V_{Ma}(m_\bullet), M_a = m_a - \Delta) \approx_{law} v(m_\Delta). \tag{23}$$

In the case $(2\alpha \leq \beta, n = 1)$, the limit distribution is also independent of the AM/DM cluster type but has an infinite mean and generation function

$$Ez^{V_{Ma}(m_\bullet)} = \phi_F(-e^{\beta\Delta/2}(1-z)^{1/2}). \tag{24}$$

**II.** *The case* $(\alpha < \beta < 2\alpha, n = 1)$, *AM cluster*

Let $M = m_\bullet - \Delta$, then the limit $V_M(m_\bullet)$ distribution has a generation function $\phi_F(-e^{\alpha\Delta}\tilde{A}_M(z))$, where $\tilde{A}_M(z)$ is defined by the equation

$$(1-z)\int_{\tilde{A}_M(z)}^{\infty} \phi_F(-u)u^{-\beta/\alpha-1}du = I_F, \tag{25}$$

$$I_F := \int_0^\infty [\phi_F(-u) - 1 + u] u^{-\beta/\alpha - 1} dx.$$

At small $\lambda_0 = (1 - \alpha/\beta)$,

$$\phi_F(-e^{\alpha\Delta}\tilde{A}_M(z)) \approx \phi_F(-(1-z)^{\alpha/\beta} c\lambda(\Delta)), \quad c = (I_F \beta/\alpha)^{-\alpha/\beta} \quad (26)$$

**III.** *The case* $(\alpha < \beta < 2\alpha, n = 1)$, *DM cluster*

If $M = m_\bullet - \Delta$, then the limit $V_M(m_\bullet)$ distribution has generation function $\phi_F(\hat{A}_M(z))$, where $\hat{A}_M(z)$ is defined by the equation

$$\hat{A}_M(z)/\lambda_0 = \int_0^1 [\phi_F(\hat{A}_M(z)y) - 1 - \hat{A}_M(z)y]\mu(dy) + (z-1)\int_{e^{-\alpha\Delta}}^1 \phi_F(\hat{A}_M(z)y)\mu(dy) \quad (27)$$

and $\mu(dy) = \beta/\alpha y^{-\beta/\alpha - 1} dy$ and $\lambda_0 = (1 - \alpha/\beta)$.

At small $\lambda_0 = (1 - \alpha/\beta)$,

$$\phi_F(\hat{A}_M(z)) \approx \phi_F(\lambda_0(z-1)(e^{\beta\Delta} - 1)) \approx \phi_F((z-1)\lambda(\Delta)F_1(\Delta)), \quad (28)$$

that is

$$V_{M_\bullet}(m_\bullet) \approx_{law} \nu(m_\Delta), \quad \lambda(m_\Delta) = \lambda_0(e^{\Delta\beta} - 1).$$

**Remark 1.** The generating function $\phi_F(-(1-z)^\gamma \lambda_\Delta), 0 < \gamma < 1$ in (24, 26) corresponds to the random variable $V := \sum_{i=1}^{\nu(m_\Delta)} \varsigma_i$ with independent components $\{\nu(m_\Delta), \varsigma_1, ... \varsigma_i, ...\}$ such that $Ez^{\varsigma_i} = 1 - (1-z)^\gamma$ and $E\nu(m_\Delta) = \lambda_\Delta$.

Applying to *V* the general theory of sub-exponential distributions (see, e.g., [Denisov et al., 2010]), we get the following consequence:

if $\nu(m_\Delta)$ has all moments (it is the case of F=P/G ), then for large *n*

$$P(V = n) \approx \lambda_\Delta P(\varsigma_1 = n) = \lambda_\Delta \gamma n^{-1-\gamma}/\Gamma(1-\gamma)(1 + o(1)), \quad (29)$$

where $\Gamma(x)$ is the Euler Gamma function.

For the regular ETAS model, Saichev et al. (2005) obtained a similar result for the total number of cluster events *V*. Namely, at the critical regime $P(V = n) = O(n^{-1-\gamma})$, where $\gamma = \max(1/2, \alpha/\beta)$ is the same as in (24, 26). This result contains no constraints on the proximity of $\alpha$ and $\beta$.

**Remark 2.** According to (22), the function $M \to P\{V_M(m_\bullet) = 0\}$ determine the magnitude distribution of the strongest aftershock $\mu_a : P\{\mu_a < M\} = Ez^{V_M(m_\bullet)}\big|_{z=0}$. In the limit case this connection requires additional verification. In regimes (n<1) and $(\alpha < \beta < 2\alpha, n = 1)$, it does not cause difficulties because of the simplicity of the limit generating function of $V_{M_a}(m_\bullet)$. The case $(2\alpha > \beta, n = 1)$ requires additional analytics, which are presented in the Appendix (see A39, A55-A57).

**Comments**

**1**. *Relationship between $V_{M_a}(m_\bullet)$ and $\nu(m_\bullet)$ distributions*

In the ETAS(F) model, statistics $\{V_M(m_\bullet), M = m_\bullet - \Delta\}$ and $\nu(m_\bullet)$ cannot have the same RT type [Molchan et al, 2022]. Replacing threshold $M = m_\bullet - \Delta\}$ with $M_a = m_a - \Delta$ corrects the situation, at least in part. According to (23), in subcritical regime, the limit $V_{M_a}(m_\bullet)$ distribution coincides with the $\nu(m_\Delta)$ distribution having mean $\lambda(m_\Delta) = e^{\beta\Delta}$. This is main result for applications.

*2. AM/DM clusters*

The probabilistic structure of AM/DM clusters is similar and differs only in $(f_1(m), \lambda(m))$ characteristics that become close at $m_\bullet >> 1$. Therefore, it is quite expected that the limit $V_{M_a}(m_\bullet)$ distributions for both cluster cases will be the same. These intuitive considerations proved themselves in the subcritical regime and even in the critical one, if $\alpha = \beta$ or $2\alpha \leq \beta$. Radical differences in the $V_{M_a}(m_\bullet)$ distributions for *AM/DM* clusters arise when $(2\alpha > \beta, n = 1)$, i.e. when the variance of the number of direct aftershocks $\nu(\mu - random)$ is unbounded and the regime is critical.

*3. Correctness of the $\Delta$-analysis*

The $\Delta$-analysis is related to the $V_{M_\bullet}(m_\bullet)$ statistics. The existence of a nontrivial limit $V_{M_\bullet}(m_\bullet)$-distribution is possible only if $m_a - m_\bullet = const$. This is only true in the critical regime under the condition $(\alpha < \beta < 2\alpha)$. However, in the practically interesting case then $\alpha/\beta \sim 1$, the difference $m_a - m_\bullet$ depends weakly on $m_\bullet$ independently of the criticality index, $n \leq 1$. Therefore, in this case too, the $V_M(m_\bullet)$ distributions for both M-thresholds: $M_a, M_\bullet$ can be expected to be close to each other at least in the central range of values.

But then, according to (23, 26, 28), $V_{M_\bullet}(m_\bullet)$ distribution should have approximately the common RT type with the distribution of direct aftershocks. These qualitative findings are consistent with the empirical results of Shebalin et al (2018, 2020), where the authors found Geometric distributions for the $V_{M_\bullet}(m_\bullet)$ and $\nu(m_\bullet)$ statistics in DM clusters.

The data of Zaliapin and Ben-Zion (2016) refer to (AM) clusters with arbitrary initial magnitudes. In this case, the empirical $V_{M_\bullet}(m_\bullet)$ distribution behaves differently: roughly speaking, power-law $Cn^{-3}$ in areas with high heat flow and lognormal otherwise. This case requires additional analysis because it is influenced by two factors: misidentification of cluster triggers and control of the upper magnitude limit by high heat flow.

## 4. Conclusion.

The key elements of the presented theoretical analysis of seismic clusters are as follows:
1) instead of the traditional ETAS model, its generalization ETAS(F) with arbitrary F distribution of direct aftershocks is used;………………………………………………………
2) the choice of a special class of F distributions possessing the RT property. This class is interesting for applications and important for taking into account Utsu's law;
3) the traditional selection of strong cluster events was changed: the lower $m_\bullet - \Delta$ magnitude threshold was replaced by $m_a - \Delta$, where $m_a$ is the peak value of the distribution of the strongest aftershock.

In the subcritical regime, $V_{m_a-\Delta}(m_\bullet)$-statistic has a quite simple limit distribution identical to the $\nu(m_\Delta)$ distribution with the mean value $E\nu(m_\Delta) = e^{\beta\Delta}$. This result provides an additional opportunity to test the ETAS(F) model. Traditional statistics $V_{m_\bullet-\Delta}(m_\bullet)$ cannot have the same RT type (say, geometric distribution) as $\nu(m_\bullet)$. However, when the seismic parameter $\alpha/\beta$ is close to 1, the RT types turn out to be close for clusters with a dominant initial magnitude.
In fact, the empirical results of Shebalin et al. (2018, 2020) support the hypothesis about the geometric distribution of $V_{m_a-\Delta}(m_\bullet)$ and $\nu(m_\bullet)$. On the other hand, the $\alpha/\beta \approx 1$ relation is generally accepted although it is of a qualitative nature. Therefore these facts can be considered as a confirmation of the concept of epidemic-type clustering embedded in the ETAS(F) model. To replace qualitative conclusions with rigorous ones, it is necessary to have good estimates of $\alpha/\beta$ and to specify the condition $m_\bullet \gg 1$ that is acceptable for practice.

## Acknowledgments

We are grateful to Prof. D. Sornette for a fruitful discussion of the RT property in seismicity models.

# Appendix

## Appendix 1 Auxiliary statements

**Statement 5 [Molchan &Varini, 2023].**

Let $v^{(\lambda)}, \lambda > 0$ be a family of random variables with the distribution: $P(v^{(\lambda)} = n) = p_n(\lambda), n \geq 0$ such that $Ev^{(\lambda)} = \lambda$, $E[v^{(\lambda)}]^2 < \infty$ and $p_0(\lambda) \to 0, \lambda \to \infty$.

Assume that for any $\lambda > 0$

$$Ez^{v^{(\lambda)}} = \phi(\lambda(z-1)), |z| < 1.$$

Then

   a) $\phi(w), w < 0$ function is analytic, increasing from $\phi(-\infty) = 0$ to $\phi(0) = 1$;

   b) $0 \leq \phi'(w) < \phi'(0) = 1$; $0 \leq \phi''(w) < \phi''(0) < \infty$; (A1)

   c) $0 \leq \phi(w) - 1 - w < c(w^2 \wedge |w|)$. (A2)

**Statement 6.** [Molchan &Varini, 2023].

Consider ETAS(F) model with magnitude and productivity characteristics $(f_1(m), \lambda(m))$. Let $\varphi_F(z|m) = Ez^{v(m)}$ be the generating function for direct aftershocks of an *m*-event. Then $DM(m_\bullet)$ cluster in the ETAS(F) model can be viewed as $AM(m_\bullet)$ cluster in the ETAS($\hat{F}$) model such that the magnitude law has the form

$$\hat{f}_1(m) = f_1(m) / F_1(m_\bullet), 0 \leq m \leq m_\bullet \qquad (A3)$$

and the off-spring law $\hat{F}$ is contiguous with F:

$$\hat{p}_k(m) = p_k(m) F_1^k(m_\bullet) / \varphi_F(F_1(m_\bullet)|m). \qquad (A4)$$

The $\hat{F}$ generating function looks as follows

$$\varphi_{\hat{F}}(z|m) = \varphi_F(zF(m_\bullet)|m) / \varphi_F(F(m_\bullet)|m). \qquad (A5)$$

## Appendix 2. Proofs.

**Proof of Statement 1**

Let $\varphi_n(z)$ be generating functions of integer variables $v_n$ with means $\lambda_n \uparrow \infty$ and $R_p$ is the random thinning operation with the parameter $p \in (0,1)$. Then the generating function of $R_{\lambda/\lambda_n} v_n$ is $\psi_n(z|\lambda) = \varphi_n(1 + \lambda(z-1)/\lambda_n)$.

Since $\varphi_n(w)$ is analytic and $|\varphi_n(w)| < 1$ in the disk $|w| < 1$, the same properties hold for the function $\psi_n(z|\lambda)$ in the $S_n(\lambda)$ disk with the diameter $(1 - 2\lambda_n/\lambda, 1)$.

Under these conditions, there exists a subsequence $\psi_{\tilde{n}}(z|\lambda)$ converging to an analytic function $\Psi_N(z|\lambda)$ in any disk with diameter (-N,1) (Stoilov, 1962). These limits analytically continue each other and represent one analytic function $\Psi(z|\lambda)$ in the half-plane $\operatorname{Re} z < 1$. Since $\psi_n(z|\lambda) = \psi_n(1+\lambda/\mu(z-1)|\mu)$ on $S_n(\lambda \vee \mu)$, the same is true for the limit function $\Psi(z|\lambda)$
Hence
$$\Psi(z|\lambda) = \Psi(1+\lambda(z-1)|1).$$
The convergence of the generating functions $\psi_{\tilde{n}}(z|\lambda)$ in the disk $|z|<1$ entails convergence $R_{\lambda/\lambda_n} \nu_{\tilde{n}}$ in distribution to an integer variable $\nu(\lambda)$. By virtue of Fatou's theorem, $E\nu(\lambda) = \lambda$ since $ER_{\lambda/\lambda_{\tilde{n}}} \nu_{\tilde{n}} = \lambda$. Hence, the $\nu(\lambda)$ set has the RT property, since $Ez^{\nu(\lambda)} = \phi(\lambda(z-1))$ and $E\nu(\lambda) = \lambda$.

**Proof of Statement 4**

*A. The basic equations for AM clusters*

Let $Ez^{\nu(m)} = \phi(\lambda(m)(z-1))$, M is a fixed magnitude threshold,
$$U_M(z|m) = Ez^{V_M(m)} \text{ and } U_M(z) = \int_0^\infty U_M(z|m)f_1(m)dm. \tag{A6}$$

Using the main equation (20), we have
$$U_M(z|m) = \sum_0^\infty P(\nu(m)=n)[Ez^{V_M(m)+1_{m>M}}]^n$$
$$= \phi(\lambda(m)([\int_0^M U_M(z|x)f_1(x)dx + z\int_M^\infty U_M(z|x)f_1(x)dx] - 1))$$
$$= \phi(\lambda(m)[\int_0^\infty (U_M(z|x)-1)f_1(x)dx + (z-1)\int_M^\infty (U_M(z|x)-1)f_1(x)dx + (z-1)\overline{F}_1(M)]). \tag{A7}$$

Now we fix the initial magnitude $m_\bullet$ and $M=M(m_\bullet)$. Consider
$$a_u(z) = \lambda(m_\bullet)\int_u^\infty (U_M(z|x)-1)f_1(x)dx, u \geq 0. \tag{A8}$$

Note that $a_u(z) \geq a_0(z)$ since $U_M(z|m) \leq 1$. We can rewrite (A7) as follows
$$U_M(z|m) = \phi(\lambda(m)\lambda^{-1}(m_\bullet)A_M(z)), \tag{A9}$$
where
$$A_M(z) := [a_0(z) + (z-1)a_M(z) + (z-1)\lambda(m_\bullet)\overline{F}_1(M)]. \tag{A10}$$

Let us subtract 1 from both parts of equality (A8) and integrate them over interval $(u,\infty)$ with weight $\lambda(m_0)f_1(m)$. We obtain a closed system of two equations with respect to $a_u(z), u = 0, M$:
$$a_u(z) = \int_u^\infty \{\phi(\lambda(m)/\lambda(m_\bullet)A_M(z)) - 1\}\lambda(m_\bullet)f_1(m)dm. \tag{A11}$$

.By (A9), the solution of our problem is the limit of

$$U_M(z|m_\bullet) = \phi(A_M(z)), \quad M = M(m_0), m_0 \gg 1. \tag{A12}$$

If a nontrivial limit of $U_M(z|m_\bullet)$ exists, the limit $A_M(z)$ function is separated from 0 and $\{-\infty\}$, where $\phi(\cdot)$ is 1 and 0 respectively. So we will search for a solution under the assumption that at fixed $z \in (0,1)$

$$-A_- \leq A_M(z) \leq -A_+ \quad , \tag{A13}$$

Let's substitute

$$\lambda(m) = \lambda_0 e^{\alpha \cdot m} \quad \text{and} \quad f_1(m) = \beta e^{-\beta m}/(1-e^{-\beta M_1}), 0 \leq m \leq M_1 = K_1 m_\bullet$$

into (A11) and change the variables under the integrals: $m \to x = \exp(\beta(m_\bullet - m))$. Omitting the $(1-e^{-\beta M_1}) = 1 + o(1)$ term we have

$$a_u(z) = \lambda_0 \int_\delta^{\exp(\beta(m_\bullet - u))} \{\phi_F(x^{-\alpha/\beta} A_M(z)) - 1\} dx\, e^{-(\beta-\alpha)m_\bullet}, \tag{A14}$$

where

$$\delta = \exp(\beta(m_\bullet - M_1)) = \begin{cases} 0 & , M_1 = \infty \\ e^{-\beta(K_1 - 1)m_\bullet} & , M_1 = K_1 m_\bullet \end{cases}.$$

We have $y(x) = A_M(z) x^{-\alpha/\beta} \to 0$ as $x \to \infty$ by virtue of (A13), and $\phi(y) - 1 = y(1 + o(1)), y \to 0$ by virtue of Statement 5. Appling the L'Hôpital's rule, we have for large R

$$\int_1^R [\phi_F(y(x)) - 1]dx = \int_1^R y(x)dx(1+o(1)) \approx \begin{cases} A_M(z) R^{1-\alpha/\beta}/(1-\alpha/\beta) & , \alpha < \beta \\ A_M(z) \ln R & , \alpha = \beta \end{cases}. \tag{A15}$$

In addition, since $0 \leq \phi_F \leq 1$, we have

$$\int_\delta^1 |\phi_F(y(x)) - 1| dx \leq 1 \tag{A16}$$

Relations (A15, A16) are key to simplify (A14) when $m_\bullet$ and $M = m_a - \Delta$ are large.

*A1. The case* $(\alpha < \beta, n < 1)$

In this case

$$M = (\alpha/\beta)m_\bullet + \beta^{-1} \ln(\lambda_0/(1-n)) - \Delta \tag{A17}$$

and $R = e^{\beta(m_\bullet - u)} \gg 1$ both for u=0 and u=M. Therefore (A15) give

$$a_0(z) \approx \lambda_0/(1-\alpha/\beta) \cdot A_M(z) = n A_M(z), \tag{A18}$$

$$a_M(z) = A_M(z) \cdot O(e^{-(1-\alpha/\beta)\alpha m_\bullet}). \tag{A19}$$

Substituting (A18, A19) in (A10), we have

$$A_M(z) \approx n \cdot A_M(z) + A_M(z) o(1)(z-1) + (z-1)(1-n)e^{\beta\Delta}.$$

That is $A_M(z) \approx (z-1)e^{\beta\Delta}$. As a result we get

$$U_M(z|m_\bullet) \Rightarrow \phi_F((z-1)e^{\beta\Delta}) . \tag{A20}$$

**A2. The case** $(\alpha = \beta, n < 1)$

In this case $n = \lambda_0 \beta M_1 = \lambda_0 \beta m_\bullet K_1$ and

$$M = m_\bullet + \beta^{-1} \ln(\lambda_0/(1 - n/K_1)) - \Delta . \tag{A21}$$

Therefore $R = e^{\beta(m_\bullet - u)} \gg 1$ both for u=0 and u=M since $\lambda_0 m_\bullet = const$. Using (A10), we have

$$a_0(z) \approx \lambda_0 \beta m_\bullet A_M(z) = n/K_1 A_M(z) , \tag{A22}$$

$$a_M(z) \approx \lambda_0 \ln[(1 - n/K_1)e^{\beta\Delta}/\lambda_0] A_M(z) = O(\ln m_\bullet / m_\bullet) A_M(z) , \tag{A23}$$

and by (A10),

$$A_M(z) \approx n/K_1 \cdot A_M(z) + A_M(z)o(1)(z-1) + (z-1)(1 - n/K_1)e^{\beta\Delta} .$$

Hence, $U_M(z|m_\bullet)$ limit is again (A20).

**A3. The case** $(2\alpha \leq \beta, n = 1)$

In this case

$$M = 2\alpha\beta^{-1}m_\bullet - \beta^{-1}\ln m_\bullet \cdot 1_{\beta=2\alpha} + \beta^{-1}\ln B - \Delta , \tag{A24}$$

where $B = 2(\beta\phi_F''(0))^{-1}(\beta - 2\alpha + 1_{\beta=2\alpha})$. Since $R = \beta(m_\bullet - M) \gg 1$, equation (A14) for $a_M(z)$ is analyzed as above using (A15) asymptotics. We have

$$a_M(z) \approx A_M(z)e^{-\alpha m_\bullet}[Ce^{-\alpha(1-2\alpha/\beta)m_\bullet}1_{2\alpha<\beta} + C_1 m_\bullet^{1/2} 1_{2\alpha=\beta}]$$

$$= A_M(z)e^{-\alpha m_\bullet}[o(1)1_{2\alpha<\beta} + o(m_\bullet)1_{2\alpha=\beta}] . \tag{A25}$$

Now we consider the equation for $a_0(z)$. Since

$$n = \lambda_0/(1 - \alpha/\beta) = \int_0^\infty \lambda(m)f_1(m)dm = 1,$$

we have

$$a_0(z) = \int_\delta^{\exp(\beta m_\bullet)} [\phi_F(A_M(z)x^{-\alpha/\beta}) - 1 - A_M(z)x^{-\alpha/\beta}]dx \times \lambda_0 e^{-(\beta-\alpha)m_\bullet} + A_M(z) . \tag{A26}$$

To simplify (A26), we will need the properties of $\phi_F(w)$ from Statement 5:

$$0 \leq \phi_F(w) - 1 - w < c(w^2 \wedge |w|) , \tag{A27}$$

$$\phi_F(w) - 1 - w = \ddot\phi_F(0)w^2/2(1 + o(1)) . \tag{A28}$$

Acting in the same way as in (A16, A17) we have

$$\int_1^R [\phi_F(A_M(z)x^{-\alpha/\beta}) - 1 - A_M(z)x^{-\alpha/\beta}]dx \approx \int_1^R [0.5\ddot\phi_F(0)A_M^2(z)x^{-2\alpha/\beta}dx(1+o(1))$$

$$= 1/2\ddot\phi(0)A_M^2(z)[R^{1-2\alpha/\beta}/(1 - 2\alpha/\beta) \cdot 1_{2\alpha\neq\beta} + 1_{2\alpha=\beta}\ln R](1 + o(1)), \qquad R \gg 1 \tag{A29}$$

$$\int_{\delta}^{1} \left|\phi_F(A_M(z)x^{-\alpha/\beta}) - 1 - A_M(z)x^{-\alpha/\beta}\right|dx \le c\int_0^1 |A_M^2(z)|x^{-\alpha/\beta}dx < C_z. \tag{A30}$$

Substituting $R = e^{\beta m_\bullet}$, equation (A26) is transformed by (A29) as follows

$$(\ddot{\phi}_F(0)/2)A_M^2(z)\beta[(\beta - 2\alpha)^{-1}1_{2\alpha \ne \beta} + m_\bullet 1_{2\alpha = \beta}]e^{-\alpha m_\bullet} \approx a_0(z) - A_M(z)$$

$$= (1-z)a_M(z) + (1-z)\lambda(m_\bullet)\overline{F}_1(M)$$

$$= e^{-\alpha m_\bullet}(o(1) + o(m_\bullet)1_{2\alpha = \beta}) + (1-z)\lambda_0 e^{\alpha m_\bullet - \beta M}. \tag{A31}$$

Substituting here $M$ from (A24), we obtain $A_M^2(z) = (1-z)e^{\beta\Delta}$. Therefore,

$$\lim U_M(z|m_\bullet) = \phi_F(-(1-z)^{1/2}e^{\beta\Delta/2}), 2\alpha \le \beta, n = 1. \tag{A32}$$

### A4. The case $(\alpha < \beta < 2\alpha, n = 1)$

We have

$$M = m_\bullet + \alpha^{-1}\ln\psi(\lambda_0) - \Delta := m_\bullet - \Lambda, \tag{A33}$$

where $\psi(\lambda_0)$ is given in (17). Hence, by (A14) and (A27, A28)

$$a_M(z) = \int_{\delta}^{\exp(\beta\Lambda))} \{\phi_F(x^{-\alpha/\beta}A_M(z)) - 1\}dx \cdot \lambda_0 e^{-(\beta-\alpha)m_\bullet}$$

$$\approx [(\beta/\alpha)|A_M(z)|^{\beta/\alpha}\int_{|A_M(z)|e^{-\alpha\Lambda}}^{\infty} \phi_F(-u)u^{-\beta/\alpha-1}du - e^{\beta\Lambda}]\lambda_0 e^{-(\beta-\alpha)m_\bullet}. \tag{A34}$$

For $a_0(z)$ we use equation (A26) to have

$$a_0(z) = \int_{\delta}^{\exp(\beta m_\bullet)}[\phi_F(A_M(z)x^{-\alpha/\beta}) - 1 - A_M(z)x^{-\alpha/\beta}]dx \times \lambda_0 e^{-(\beta-\alpha)m_\bullet} + A_M(z). \tag{A35}$$

Due to (A27, A28) the following integral

$$I_F := \int_0^{\infty}[\phi_F(-u) - 1 + u]u^{-\beta/\alpha-1}dx \tag{A36}$$

is finite if $(\alpha < \beta < 2\alpha$. Therefore, in the limit the integration interval in (A28) can be replaced by a semi-straight line. Then, (A34) and (A235) give

$$(\beta/\alpha)|A_M(z)|^{\beta/\alpha}I_F\lambda_0 e^{-(\beta-\alpha)m_\bullet} \approx (1-z)a_M(z) + (1-z)\lambda_0 e^{\beta\Lambda}e^{-(\beta-\alpha)m_\bullet}$$

$$= (\beta/\alpha)|A_M(z)|^{\beta/\alpha}(1-z)\int_{|A_M(z)|e^{-\alpha\Lambda}}^{\infty}\phi_F(-u)u^{-\beta/\alpha-1}du \cdot \lambda_0 e^{-(\beta-\alpha)m_\bullet}.$$

Finally,

$$(1-z)\int_{|A_M(z)|e^{-\alpha\Lambda}}^{\infty}\phi_F(-u)u^{-\beta/\alpha-1}du = I_F. \tag{A37}$$

Here $\exp(-\alpha\Lambda) = \psi(\lambda_0)e^{-\alpha\Delta}$. The obtained equation can be simplified at small $\lambda_0 = (1 - \alpha/\beta)n$. In this case $\psi(\lambda_0) = \lambda_0(1 + o(1))$ and

$$\int_{\varepsilon}^{\infty}\phi_F(-u)u^{-\beta/\alpha-1}du = \alpha\beta^{-1}\varepsilon^{-\beta/\alpha}(1+o(1)) \text{ as } \varepsilon \to 0.$$

Hence at small $\lambda_0$,

$$A_M(z) \approx -\lambda_0 e^{\alpha\Delta}(I_F\beta/\alpha)^{-\alpha/\beta}(1-z)^{\alpha/\beta}, \qquad (A38)$$

that is

$$U_M(z|m_\bullet) \approx \phi_F(-(1-z)^{\alpha/\beta}ce^{\alpha\Delta}), \qquad c = \lambda_0(I_F\beta/\alpha)^{-\alpha/\beta}.$$

**The relation (A37) at z=0**. This case is equivalent to (17, 18), i.e.

$$\psi + (\beta/\alpha - 1)\psi^{\beta/\alpha}\int_0^\psi u^{-\beta/\alpha-1}[\phi_F(-u)-1+u]du = \lambda_0, \qquad (A39)$$

where $\psi = |A_M(z=1)|e^{-\alpha\Delta}$. To verify this, it is enough to substitute $I_F$ from (A36) into (A37). After simple algebra we obtain at z=0

$$\int_0^\psi (\phi_F(-u)-1+u)u^{-\beta/\alpha-1}du = \int_\psi^\infty (1-u)u^{-\beta/\alpha-1}du = \alpha/\beta \cdot \psi^{-\beta/\alpha} - \psi^{1-\beta/\alpha}/(\beta/\alpha-1).$$

To get (A39) we should multiply the obtained ratio by $\psi^{\beta/\alpha}(\beta/\alpha-1)$ and take into account that $\lambda_0 = 1 - \alpha/\beta$.

**B.DM clusters**

Now we consider $DM(m_\bullet)$ cluster in ETAS(F) model with RT property of F. According to Statement 6, statistics $\nu(m)$ in such cluster has the following generating function

$$Ez^{\nu(m)} = \phi_F(\lambda(m)(zF_1(m_\bullet)-1))/\phi_F(\lambda(m)(F_1(m_\bullet)-1)). \qquad (A40)$$

Following (A7-A11) we can obtain the equation for

$$U_M(z|m) = Ez^{V_M(m)} \text{ and } U_M(z) = \int_0^{m_\bullet}U_M(z|m)\hat{f}_1(m)dm,$$

where $\hat{f}_1(m) = f_1(m)/F_1(m_\bullet), 0 \leq m \leq m_\bullet$.

Using notation

$$\hat{a}_u(z) = \lambda(m_\bullet)\int_u^{m_\bullet}(U_M(z|m)-1)f_1(m)dm,$$

one has

$$U_M(z|m) = \phi_F(\lambda(m)\lambda^{-1}(m_\bullet)\hat{A}_M(z))/\phi_F(-\lambda(m)\overline{F}_1(m_\bullet)), \qquad (A41)$$

where

$$\hat{A}_M(z) := \hat{a}_0(z) + (z-1)\hat{a}_M(z) + (z-1)\lambda(m_\bullet)\overline{F}_1(M) - z\lambda(m_\bullet)\overline{F}_1(m_\bullet). \qquad (A42)$$

Let's subtract 1 from both parts of equality (A41) and integrate over interval $(u, m_\bullet)$ with weight $\lambda(m_\bullet)f_1(m)$. Then we obtain a closed system of two equations with respect to $\hat{a}_u(z), u = 0, M$ :

$$\hat{a}_u(z) = \int_u^{m_\bullet}\{\phi_F(\lambda(m)/\lambda(m_\bullet)\hat{A}_M(z))/\phi_F(-\lambda(m)\overline{F}_1(m_\bullet)) - 1\}\lambda(m_\bullet)f_1(m)dm. \qquad (A43)$$

After replacement of variables $m \to x = \exp(\beta(m_\bullet - m))$ we obtain the analogue of relation (A14)

$$\hat{a}_u(z) = \int_1^{\exp(\beta(m_\bullet - u))} \{\phi_F(x^{-\alpha/\beta}\hat{A}_M(z))/\phi_F(-x^{-\alpha/\beta}\varepsilon) - 1\}dx \cdot \varepsilon \quad , \tag{A44}$$

where $\varepsilon = \lambda_0 e^{-(\beta-\alpha)m_\bullet}$.

In the case $\alpha = \beta$, $\lambda_0 m_\bullet = const$. Therefore $\varepsilon = o(1)$, when $\alpha \le \beta$ and $m_\bullet \gg 1$.

Our goal is the limit function for $\quad U_M(z|m_\bullet) = \phi_F(\hat{A}_M(z))/\phi_F(-\varepsilon)$.

Since $1 - x < \phi_F(-x) \le 1$,

$$\lim U_M(z|m_\bullet) = \phi_F(\lim \hat{A}_M(z)). \tag{A45}$$

Hence, as above, we can assume that at fixed $z \in (0,1)$

$$-A_- \le \hat{A}_M(z) \le -A_+$$

To simplify (A43), note that.

$$1 - \phi_F(-x) \ge 0 \text{ and } 1 - \varepsilon \le \phi_F(-x^{-\alpha/\beta}\varepsilon) \le 1, x \ge 1.$$

From here

$$\hat{a}_u(z) = \{\varepsilon\int_1^{\exp(\beta(m_\bullet - u))}[\phi_F(x^{-\alpha/\beta}\hat{A}_M(z)) - 1]dx \cdot + Q_\varepsilon(u)\}(1 + O(\varepsilon)), \tag{A46}$$

where $u \le m_\bullet$ and

$$Q_\varepsilon(u) = \varepsilon\int_1^{\exp(\beta(m_\bullet - u))}[1 - \phi_F(-x^{-\alpha/\beta}\varepsilon)]dx \cdot \le \int_1^{\exp(\beta(m_\bullet - u))} x^{-\alpha/\beta}\varepsilon^2 dx.$$

Therefore

$$Q_\varepsilon(u) \le \varepsilon^2 \cdot \begin{cases}[e^{(\beta-\alpha)(m_\bullet - u)} - 1]/(1 - \alpha/\beta) \\ \beta(m_\bullet - u)\end{cases} < \varepsilon\begin{cases}ne^{-(\beta-\alpha)u} & \alpha < \beta \\ C & \alpha = \beta\end{cases}. \tag{A47}$$

Relations (A46, A47) allow us to transfer the proofs of the results for AM clusters to DM clusters in the following regimes: $(\alpha \le \beta, n \le 1) \cup (2\alpha \le \beta, n = 1)$, where $M$ is the same for these clusters.

**B2. .The case $(\alpha < \beta < 2\alpha, n = 1)$, DM cluster**

Let's remember that $M = m_\bullet - \Delta$, $\lambda_0 = 1 - \alpha/\beta$, $\varepsilon = \lambda_0 e^{-(\beta-\alpha)m_\bullet}$.

Replacing in (A44) the variable $x$ by $y = x^{-\alpha/\beta}$ we have

$$\hat{a}_u(z) = \int_{\exp(-\alpha(m_\bullet - u))}^1 \{\phi_F(y\hat{A}_M(z))/\phi_F(-y\varepsilon) - 1\}\mu(dy) \cdot \varepsilon, \tag{A48}$$

where $\mu(dy) = \beta/\alpha y^{-\beta/\alpha - 1}dy$.

By (A42),

$$\hat{A}_M(z) = \hat{a}_0(z) + (z-1)\hat{a}_M(z) + \varepsilon(z-1)e^{\beta \cdot \Delta} - z\varepsilon \tag{A49}$$

By (A48),

$$\hat{a}_0(z) = \int_{\exp(-\alpha m_\bullet)}^1 \{\phi_F(y\hat{A}_M(z)) - 1 - y\hat{A}_M(z)\}/\phi_F(-y\varepsilon) \cdot \mu(dy) \cdot \varepsilon - I_2 + I_3, \qquad (A50)$$

where

$$I_2 = \int_{\exp(-\alpha m_\bullet)}^1 \{\phi_F(y\varepsilon) - 1 - y\varepsilon\}/\phi_F(-y\varepsilon)\mu(dy) \cdot \varepsilon = O(\varepsilon^3), \qquad (A51)$$

$$I_3 = (\hat{A}_M(z) + \varepsilon)\int_{\exp(-\alpha m_\bullet)}^1 y/\phi_F(-y\varepsilon) \cdot \mu(dy) \cdot \varepsilon = (\hat{A}_M(z) + \varepsilon)(1 - \varepsilon/\lambda_0) + O(\varepsilon^2). \qquad (A52)$$

The estimations in (A51, A52) follow from (A27). Each of the relations (A49), (A50) allows us to find an expression for $Q := (\hat{A}_M(z) - \hat{a}_0(z))/\varepsilon + 1$.

According to (A48),

$$Q = (z-1)\int_{\exp(-\alpha\Delta)}^1 \phi_F(y\hat{A}_M(z))/\phi_F(-y\varepsilon)\mu(dy)$$

Given (A50), we have

$$Q = -\int_{\exp(-\alpha m_\bullet)}^1 \{\phi_F(y\hat{A}_M(z)) - 1 - y\hat{A}_M(z)\}/\phi_F(-y\varepsilon) \cdot \mu(dy) \cdot + \hat{A}_M(z)/\lambda_0 + o(1).$$

Since $I_F := \int_0^\infty [\phi_F(-u) - 1 + u]u^{-\beta/\alpha - 1}dx$ is bounded and $\phi_F(-y\varepsilon) - 1 = O(\varepsilon)$ we can go to the limit to get the equation for the limit $\hat{A}_M(z)$ function:

$$\hat{A}_M(z)/\lambda_0 = \int_0^1 [\phi_F(\hat{A}_M(z)y) - 1 - \hat{A}_M(z)y]\mu(dy) + (z-1)\int_{e^{-\alpha\Lambda}}^1 \phi_F(\hat{A}_M(z)y)\mu(dy) \qquad (A53)$$

Finally, the desired generating function of $V_M(m_\bullet)$ is $\phi_F(\hat{A}_M(z))$.

**The case of small $\lambda_0$**. When $\lambda_0$ is small, equation (A53) can be simplified. Let's substitute $\hat{A}_M(z)/\lambda_0 = B(z)$ into (A53), then we obtain

$$B(z) = \int_0^1 [\phi_F(B(z)\lambda_0 y) - 1 - B(z)\lambda_0 y]\mu(dy) + (z-1)\int_{e^{-\alpha\Lambda}}^1 \phi_F(B(z)\lambda_0 y)\mu(dy).$$

For $\lambda_0 \ll 1$

$$B(z) \approx 0.5\ddot{\phi}(0)B^2(z)\lambda_0^2 \int_0^1 y^2\mu(dy) + (z-1)\int_{e^{-\alpha\Delta}}^1 (1 + B(z)\lambda_0 y)\mu(dy),$$

that is

$$B(z) \approx O(\lambda_0^2) + (z-1)(e^{\beta\Delta} - 1) + (z-1)B(z)\beta\Delta\lambda_0.$$

Hence

$$\hat{A}_M(z) = \lambda_0(z-1)(e^{\beta\Delta} - 1)(1 + o(1)). \qquad (A54)$$

**The relation (A53) at z=0**. This case is equivalent to (17), i.e.

$$\psi + (\beta/\alpha - 1)\psi^{\beta/\alpha}\int_0^\psi u^{-\beta/\alpha - 1}[\phi_F(-u) - 1 + u]du = \lambda_0(1 - e^{-\beta\Delta}), \qquad (A55)$$

where $\psi = |\hat{A}_M(z=0)|e^{-\alpha\Delta}$. To see this, let us split the integration interval (0,1) in (A53) into two $(0,\rho) \cup (\rho,1)$, where $\rho = e^{-\alpha\Delta}$. Then (A53) at z=0 is as follows

$$-|A|/\lambda_0 = \int_0^\psi [\phi_F(-u) - 1 + u]\mu(du)|A|^{\beta/\alpha} - \int_\rho^1 (1-|A|y)\mu(dy) \ , \tag{A56}$$

where $A = \hat{A}_M(z=1)$ and

$$\int_\rho^1 (1-|A|y)\mu(dy) = \rho^{-\beta/\alpha} - 1 - |A|(\rho^{-\beta/\alpha+1} - 1)/\lambda_0 \ . \tag{A57}$$

Let us substitute (A57) into (A56) and multiply both parts of the equality (A56) by $\lambda_0 \rho^{\beta/\alpha}$. As a result, we obtain (A55). From here, for small $\lambda_0 = 1 - \alpha/\beta$

$$\hat{A}_M(z=0) \approx -\lambda_0(1 - e^{-\beta\Delta})e^{\alpha\Delta}$$

which is consistent with (A54) at z=0.